\documentclass[prb,aps,preprint]{revtex4}
\usepackage{graphicx}
\begin{document}

\title{Theory of resonant inelastic X-ray scattering in vanadium oxides: how to detect $d$-$d$ excitations?}

\author{V. Yushankhai$^{1,2}$ and L. Siurakshina$^{1,2}$}

\affiliation{ $^1$Joint Institute for Nuclear Research, 141980 Dubna, Russia\\
$^2$Max-Planck-Institut f\"ur Physik Komplexer Systeme, N\"othnitzer Stra\ss e 38,
             D-01187 Dresden, Germany\\}

\date{\today}

%%%%%%%%%%%%%%%%%---  Abstract ---%%%%%%%%%%%%%%%%%%

\begin{abstract}
Generic low-energy spectral features related to $d$-$d$ excitations in nearly cubic vanadium
perovskites are predicted for the expected $L$-edge resonant inelastic X-ray scattering (RIXS)
measurements. Model Hamiltonian describing local electronic properties, including   crystal-field
effects, of vanadium $3d$ orbitals in the basic $t_{2g}^2$ configuration is formulated with the
help of complementary {\it ab initio} quantum-chemical cluster calculations. In the presence of
$2p$-core hole, the local Hamiltonian includes strong $2p$-$3d$ electron interactions. As a
prerequisite for evaluating RIXS transition amplitudes beyond the fast collision approximation, a
symmetry-group approach is applied to generate a basis set of many-electron wavefunctions of the
intermediate core-hole states accessible in RIXS processes. Although a comprehensive description of
the core-hole multiplets still remains  a formidable task and requires using specially designed
numerical codes, for particular resonant states the analysis is simplified and the calculation of
RIXS amplitude can be carried out analytically to the end.
\end{abstract}


\maketitle

%%%%%%%%%%%%%%%%--- I ----%%%%%%%%%%%%%%%%%%%%%%%%%%%

\section{Introduction}
Soft X-ray resonant inelastic scattering (RIXS) is increasingly important in the study of
electronic properties of strongly correlated $3d$ systems.\cite{Ament11} Measurements of $d$-$d$
excitations that correspond to local rearrangements of $3d$ electrons provide a valuable piece of
information on the strong coupling between spin, orbital and lattice degrees of freedom in a
system. Last years, the advances achieved on the instrumental side of RIXS make it possible to
measure electronic excitations with energy resolution better that 0.1eV and, hence, to resolve
fine features of electronic structure not accessible previously.

The family of vanadium oxides  RVO$_3$ (with R=Y or trivalent rare earth ion) with  $t_{2g}^2$
configuration of V$^{3+}$ ions have attracted much attention because of its rich and complex
behavior.\cite{Miyasaka03,Miyasaka06,Reehuis06} In these Mott-Hubbard insulating materials with a
weak deviation from the ideal cubic perovskite structure the anticipated crystal-field (CF)
splitting of $t_{2g}$ orbitals is rather small $\sim 0.1$eV and varies with structural changes. In
these circumstances, to describe  the low-energy properties of RVO$_3$ the orbital and spin degrees
of freedom have to be treated on equal footing,  which leads to a spin-orbital superexchange
lattice model derived and studied in a series of papers.\cite{Oles05,Oles07,Horsch08,Oles12}  The
model was shown to predict diverse behavior of the system in dependence on the model parameter
chosen and complementary theoretical assumptions made. One of the basic assumptions concerns the
choice of a point symmetry at V ion, the CF parameters and their variation along the RVO$_3$ series
irrespective of any orbital and spin ordering. We note that CF splitting of $t_{2g}$ orbitals and
the  multiplet structure of $t_{2g}^2$ configuration of V ion  can be directly inferred from the
low-energy part of high-resolution RIXS spectra provided the spectral features associated with
$d$-$d$ excitations are properly identified as  bound excitonic states, i.e., the low-energy in-gap
electronic excitations produced by RIXS at the final stage in an insulating strongly correlated
system.

In the present study, we examine theoretically the spectrum of $d$-$d$ excitations with special
attention to a spin and orbital disordered phase of RVO$_3$. In Sec.II, we begin with a brief
description of several basic properties of nearly cubic vanadium perovskites,  including the
calculated electronic multiplet structure of the V$^{3+}$ ion related to the expected RIXS spectra.
Fundamentals of the vanadium $L$-edge (2$p$-3$d$ core-hole excitation) RIXS transitions are
discussed in Sec.III. Since a comprehensive analysis of relevant RIXS transitions requires
knowledge of intermediate core-hole states, this item is considered in Sec.IV in many detail. A
generic procedure for calculating the RIXS transition amplitudes together with a more detailed
consideration of the lowest-energy $d$-$d$ excitation are present in Sec.V. Summary, conclusions,
and outlook follow in Sec.VI.

\section{Vanadium oxides with nearly cubic perovskite structure}
 At room temperature the crystal lattice of
RVO$_3$ is described by an orthorhombic space group with the lattice constants $a\approx b \approx
c/\sqrt 2$. A relatively weak deviation from the ideal cubic structure is caused by a regular
rotation, tilting and distortion of VO$_6$ octahedra. As a result, the cubic degeneracy of $t_{2g}$
orbitals occupied by two electrons is removed and the accordingly weak CF splitting of $\sim 0.1$eV
is expected, which is controlled by a variation of the R-site ionic radius $r_R$ along the RVO$_3$
series.

  Interplay between weakly split $t_{2g}$ orbitals, spins and lattice degrees of freedom is in the
  hart of a rich behavior observed in RVO$_3$ family, including two types of the orbital-ordering (OO) and
  spin-ordering (SO) patterns observed and reported in a number of works, as for instance, reviewed
  shortly in Ref.\onlinecite{Reehuis11}.  An ordering, called the $C$-OO, corresponds to
  the orbital arrangement of antiferro-type in $ab-$plane and ferro-type along the $c-$axis. The
  $C$-OO together with the antiferromagnetic $G$-type spin ordering ($G$-SO, with spins
  staggered in all three directions)  occur at low temperatures, $T<$100K, in RVO$_3$
  with smallest $r_R$. The other ordering pattern is the $G$-OO accompanied with $C$-SO and
  observed at intermediate temperatures. The $G$-OO/$C$-SO is the only ordered phase in RVO$_3$
  with larger values of $r_R$. The global spin-orbital phase diagram for the  RVO$_3$ family including
  the concomitant structural phase transitions derived from the single-crystal measurements was
  reported.\cite{Miyasaka03, Reehuis11}.
Based on  a spin-orbital superexchange  model for RVO$_3$  a scenario of the puzzling transition
between the two above-mention ordered phases was suggested.\cite{Oles05,Oles07,Horsch08}

In the present paper, our main concern is to the local properties of  vanadium valence 3$d$
electrons. The local part $H_{loc}=H_{cf}+H_{d,C}$ of the underlying Hubbard-type model contains
the crystal field $H_{cf}$ and the Coulomb $H_{d,C}$ terms:
\begin{eqnarray}
\mathcal H_{cf} &=& \sum\limits_{m,\sigma}\ \epsilon^{\,d}_m\
 d_{m\sigma}^{\dagger}d_{m\sigma}, \nonumber \\
H_{d,C}&=& U \sum\limits_{m} n_{m\uparrow}n_{m\downarrow}+ \sum\limits_{m < m^{\prime}}
\left(U-\frac{5}{2}J_{m m^{\prime}}\right)n_{m}n_{m'}-
 2 \sum\limits_{m\langle m^{\prime}}J_{m m^{\prime}}{\bf S}_m {\bf S}_{m'} + \nonumber \\
&& \sum\limits_{m \not= m^{\prime}}J_{m m^{\prime}}d_{m\uparrow}^{\dagger}d_{m\downarrow}^{\dagger}
  d_{m^{\prime}\downarrow} d_{m^{\prime}\uparrow}. \label{1}
\end{eqnarray}
Here, $\epsilon^{\,d}_m$ is an energy of the $m$-th 3$d$ orbital, $U$ and $J_{m m^{\prime}}$ are
the Hubbard repulsion and  Hund's exchange parameters, respectively. When restricted only to the
$t_{2g}$ subspace, the Hamiltonian with the same exchange parameters, $J_{m m^{\prime}}=J_H$,
describes rigorously the multiplet structure of $t_{2g}^2$ and $t_{2g}^3$
configurations.\cite{Oles05,Griffith71}

Because the octahedra VO$_6$ in RVO$_3$  are tilted and distorted, the exact point symmetry at V
ion is rather low - the only operation allowed is the inversion, and the cubic 3$d$ basis
functions, $t_{2g}$ and $e_g$, are generally mixed.  Nevertheless, one expects that a higher point
symmetry, for instance, $D_{2h}$ or $D_{4h}$, can be applied  to approximate accurately the CF
levels and the electronic multiplet structure of the system. In support of this conjecture, we have
performed complementary $\textit{ab initio}$ quantum-chemical calculations\cite{Hozoi11} for a
crystal fragment containing the rotated and distorted cluster VO$_6$. The calculated valence states
are found to be hybridized  V-ion 3$d$- and oxygen 2$p$-orbitals, and in a local orthogonal frame
associated with the principal axes of a rotated octahedron these (antibonding) molecular orbitals
can be viewed as the canonical $t_{2g}$ and $e_g$ ones. In the following, they are referred to as
vanadium valence 3$d$ orbitals, and the model Hamiltonian, Eq.(\ref{1}), is applied to them as
well. To estimate model parameters, energies of several crystal-field multiplets of the valence
$d^2$ configuration, including $t_{2g}^2$ and $t_{2g}^1e_g^1$, were calculated for the cluster
VO$_6$ and compared with those derived from Eq.(\ref{1}). We found that the expected for $D_{4h}$
(or $D_{2h}$) point symmetry character of a comparatively weak non-cubic CF splitting, both within
the $t_{2g}$ and $e_g$ subspaces, is well supported by the computed spectra, which is commented in
detail below. Since the main focus is on the low-energy electronic properties and their
measurements by RIXS, the calculated multiplet structure of the $t_{2g}^2$ configuration is mostly
discussed.

To be more precise, we note that in the orthorhombic structure of RVO$_3$ there are four
non-equivalent crystallographic positions of V-ions coordinated with oxygen octahedra that are
rotated by the same angles in alternate manner. For a given V ion, assuming the $x$-, $y$-, and
$z$-axes of the local coordinate system  attached to the principal axes of the rotated oxygen
octahedron, we define the $t_{2g}$ orbitals as follows: $d_{yz}=d_{\xi}$, $d_{zx}=d_{\eta}$, and
$d_{xy}=d_{\zeta}$. The angular part of the orbital wavefunctions are described by real spherical
harmonics: $Z^2_{\xi}=(i/\sqrt{2})[Y^2_{-1}+Y^2_{1}]$, $Z^2_{\eta}=(1/\sqrt{2})[Y^2_{-1}-Y^2_{1}]$,
and $Z^2_{\zeta}=(i/\sqrt{2})[Y^2_{-2}-Y^2_{2}]$, with $Y^2_{m}$ being the canonical spherical
harmonics.

In our {\it ab initio} quantum-chemical cluster approach based on the use of MOLPRO computer
program\cite{MOLPRO}, the wave functions and energies of  electron configurations are calculated at
different levels of accuracy - from a single-reference restricted Hartree-Fock (RHF) method through
a multi-configuration self-consistent field (MCSCF) ansatz  to the multi-reference
configuration-interaction method (MRCI).\cite{Siurakshina10} The lattice fragment, whose electrons
are treated rigorously, contains the cluster VO$_6$ and eight neighboring R ions, and it is
imbedded in a large point-charge environment that simulates the Madelung potential on the cluster
ions.

In our calculations the primary emphasis was placed upon the  most widely studied compound YVO$_3$.
The structural lattice data with the actual rotation, tilting and distortion of VO$_6$,
reported\cite{Reehuis06} for YVO$_3$ were used. Although the situation is likely to depend on the
changing degree of octahedral tilting across the RVO$_3$ series\cite{Andersen07,Zhou08,Blake09}, we
expect that the generic features derived can be directly extended to other members of the RVO$_3$
family, at least to those with close values of $r_R$. Therefore, keeping in mind such an extended
usage of our cluster calculations, most representative results are reported throughout. As expected
the calculated subspaces $e_g$ and $t_{2g}$ are well separated by a large cubic CF of $\sim$1.5eV.
For the high-temperature ($T>$200K) orthorhombic structure the lowest electronic orbital is
$d_{\zeta}$, while the pair of $d_{\eta}$, $d_{\xi}$ is shifted up by $\Delta_1\sim 0.1$eV, and
these two are only weakly split by $\Delta_2\sim 10$meV. The inequality $\Delta_1\gg \Delta_2$ can
be partly explained for the high-$T$ phase by the observation\cite{Reehuis06} that in each oxygen
octahedron the tetragonal-like contraction of the V-O bonds along $z$-axis is regularly larger than
an orthorhombic distortion of the V-O bonds in the $xy$-plane. Because the energy resolution of
RIXS spectra is usually limited to $\sim$50meV, in the multiplet structure fine features  due to
orthorhombic CF level splitting of $\Delta_2\sim 10$meV can be hardly resolved. Therefore, in
subsequent theoretical analysis of the local electronic structure characteristic of the high-$T$
phase of RVO$_3$, instead of $D_{2h}$  we use the tetragonal $D_{4h}$ symmetry by setting
$\Delta_2\to 0$ and denoting the largest non-cubic CF parameter for the $t_{2g}$ manifold as
$\Delta_1=\Delta_t $. This is in contrast to the CF model suggested in Ref.\onlinecite{Oles07}.
There, the authors assumed the cubic point symmetry $O_h$ with fully degenerate $t_{2g}$ levels in
the high-$T$ phase and a symmetry reduction from $O_h$ to $D_{4h}$ (or $D_{2h}$) was associated
with structural transitions below 200K.

In high-$T$ phase of RVO$_3$ the low-energy part of the  RIXS spectra is expected to exhibit the
multiplet structure of $t_{2g}^2$ configuration. Hereafter, a simplifying approximation is used
that consists in the neglect of comparatively small alternating tilts of oxygen octahedra. Thus all
the vanadium ions are considered to be equivalent and the 3$d$ orbitals are defined in the global
($XYZ$)-coordinate system with $Z$-axis along the orthorhombic ${\bf c}$-axis, while $X$- and
$Y$-axis are rotated by 45$^{\circ}$ with respect to ${\bf a}$ and ${\bf b}$.

For the CF of $D_{4h}$ symmetry, the states of $t_{2g}^2$ configuration are sorted over the terms
$^3E$, $^3A_2$, $^1E$, $^1B_2$, $^1B_1$, and 2 $^1A_1$, where a subscript indicating the same
(even) parity of terms  is omitted for brevity. For a given term $^{2S+1}\Gamma\equiv (S\Gamma)$,
the basis functions $|t^2(S\Gamma)M\gamma\rangle$ labeled by the spin projection
$M(=-S,-S+1,...,S)$ and the basis index $\gamma$ of $\Gamma$, form a set of degenerate
eigenfunctions of the local Hamiltonian, Eq.(1). The calculated  multiplet structure is presented
schematically in Fig.1. The term energies are found to be: $\mathcal E(^3A_2)=\Delta_t$,\
$\mathcal E(^1A_1')=2J_H-\Delta\mathcal E'$,\ $\mathcal E(^1E)=2J_H$,\   $\mathcal
E(^1B_1)=\mathcal E(^1B_2)=\Delta_t+2J_H$,\   $\mathcal E(^1A_1)=5J_H+\Delta\mathcal E'$, where
$\Delta\mathcal E'\approx\Delta_t[1+(\Delta_t/J_H)]/3$, for $\Delta_t/J_H<1/2$, with the reference
energy taken at the level of lowest term $^3E$ of $t_{2g}^2$ configuration.
 From our cluster calculations, it follows also that, first, the lowest spin-triplet level
coming from the $t_{2g}^1e_g^1$ configuration falls into the same energy range - it is stuffed
between $^1B_1/^1B_2$ and $^1A_1$, but not indicated in Fig.1. Second, the estimated value of
Hund's constant is $J_H\approx 0.5$eV, which is somewhat smaller than that reported usually in
literature.\cite{Mizokawa96,Bencksier08}

 For the further purposes, it is instructive to present explicitly the basis functions of several
 terms under consideration. For instance, the ground-state term $^3E$ comprises six states:
 $|t^2(^3E)Me_1\rangle   = |d_{\eta \uparrow}d_{\zeta\uparrow}\rangle,\
 (1/\sqrt{2})[|d_{\eta \uparrow}d_{\zeta\downarrow}\rangle  +|d_{\eta
 \downarrow}d_{\zeta\uparrow}\rangle],\
 |d_{\eta \downarrow}d_{\zeta\downarrow}\rangle  $, for $\gamma = e_1$, and the other three
 states $|t^2(^3E)Me_2\rangle$ for $\gamma = e_2$ are obtained by the replacement $\eta\to\xi$.
 The first excited term  is  a triplet $|t^2(^3A_2)M\rangle   = |d_{\xi
 \uparrow}d_{\eta\uparrow}\rangle,\
 (1/\sqrt{2})[|d_{\xi\uparrow}d_{\eta\downarrow}\rangle
 +|d_{\xi\downarrow}d_{\eta\uparrow}\rangle],\
 |d_{\xi\downarrow}d_{\eta\downarrow}\rangle$, where the basis index $\gamma$ for the one-dimensional
 representation  $A_2$ is omitted.  In all sets above, states are listed according to $M=+1,0,-1$.
With respect to  RIXS, the initial (ground) and excited (final) states are denoted below as
$|t^2,[g]\rangle  $ and $|t^2,[f]\rangle  $, respectively. Here, $[g]$ and $[f]$ are the
shortenings for the quantum numbers , $[g]= (^3E)M_ge_g$ with $M_g=0,\pm 1$ and $e_g=e_{1,2}$, and
$[f]= (S_f\Gamma_f)M_f\gamma_f$.

\section{RIXS theory: basic formulation}
Consider a typical scheme of the $L$-edge RIXS measurement  for a strongly correlated $3d$-electron
system whose $N$-electron ground state is $|\Psi^N_{[g]}\rangle$. The incoming photon with momentum
$\hbar\bf k$ and energy $\hbar\omega_{\bf k}$ tuned close to the $L$-edge absorption promotes an
electron from the $2p$ shell to an empty $3d$ valence state, thus producing an intermediate
core-hole state $|(\Psi^{N+1}\underline{p})_{[i]}\rangle  $. In a subsequent radiative decay of the
core hole the emitted $(\hbar {\bf k'}, \hbar\omega_{\bf k'})$ photon leaves the $3d$ electron
system in an excited state $|\Psi^N_{[f]}\rangle  $ with momentum $\hbar(\bf k-\bf k')$ and energy
$\hbar(\omega_{\bf k}-\omega_{\bf k'})$ that are measured. For given polarization vectors
${\mbox{\boldmath $\epsilon $}}$ and ${\mbox{\boldmath $\epsilon\,'$}}$ of the incoming and and
outgoing photons, respectively, the scattering amplitude of the second-order resonant processes
governed by the transition operator ${\bf D}_{\bf k}$ can be written\cite{Ament11} as
\begin{equation}
\mathcal F^{\epsilon'\epsilon}_{fg}({\bf k'}, {\bf k}; z_k)= \langle \Phi_{[f]}^N|({\mbox{\boldmath
$\epsilon\,'$}}{\bf D}_{\bf k'})^{\dagger}\mathcal G(z_{\bf k})({\mbox{\boldmath $\epsilon$}}{\bf
D_{\bf k}})|\Phi_{[g]}^N\rangle  .
 \label{a2}
\end{equation}
Here, $\mathcal G(z_{\bf k})$ is the intermediate-state propagator which describes the system in
the presence of a core hole:

\begin{equation}
\mathcal G(z_{\bf k})=\sum\limits_{\{i\}}\frac{|(\Phi^{N+1}\underline{p})_{[i]}\rangle  \langle
(\Phi^{N+1}\underline{p})_{[i]}|}{z_{\bf k} - E_{[i]}}, \label{a3}
\end{equation}
where $z_{\bf k}=\hbar\omega_{\bf k}+i\Gamma$, for $E_{[g]}=0$, and  the lifetime broadening
$\Gamma$ of  intermediate core-hole states is taken to be independent on the state index $[i]$. We
note that $\Gamma$ depends on the $3d$-ion species probed in a measurement and, typically,
ranges\cite{Pease91,Krause79} from 0.2eV up to 0.6eV.
%The transition operator $\bf D_{\bf k}$ is defined as
%\begin{equation}
%{\bf D_{\bf k}}=\frac{1}{im\omega_{\bf k}}\sum\limits_j^{N}e^{i{\bf kr}_j}{\bf p}_j, \label{a4}
%\end{equation}
%where ${\bf r}_j$ and ${\bf p}_j$ are the position and momentum operators of electrons, and $m$ is
%the electron mass.
In the dipole limit, the transition operator $\bf D_{\bf k}$  takes the form
\begin{equation}
{\bf D_{\bf k}}=\frac{1}{\mathcal N}\sum\limits_l e^{i{\bf kR}_l}\sum\limits_{j(l)}^{N_e}({\bf r}_j
- {\bf R}_l),\label{a5}
\end{equation}
where ${\bf R}_l$ denotes the lattice sites of the  metal ions and the second summation is over
electrons belonging to the $l$-th ion.  The form Eq.(\ref{a5}) is especially suitable since we are
mainly interested in the bound excitonic states, i.e., the low-energy in-gap electronic excitations
produced by RIXS at the final stage in an insulating strongly correlated system.

 Then, in the second-quantization form the dipole operator Eq.(\ref{a5}) reads
(${\bf R}_l\equiv {\bf l}$):
\begin{equation}
{\bf D_{\bf k}}=\frac{{\mathcal R}^2}{\mathcal N}\sum\limits_{\bf l} e^{i{\bf
kl}}\left(\sum\limits_{m,\sigma}\sum\limits_{m_p}\langle Z^2_m|{\hat {\bf r}}|Y^1_{m_p}\rangle
d_{{\bf l}m\sigma}^{\dagger}p_{{\bf l}m_p\sigma}+H.c. \right).\label{a6}
\end{equation}
Here the operator $p_{{\bf l}m_p\sigma}$ destroys an electron with the orbital projection
$m_p(=0,\pm 1)$ and spin $\sigma$ in the $2p$ shell, while $d_{{\bf l}m\sigma}^{\dagger}$ creates
an extra electron with the same spin in the $m$-th orbital of the $3d$-shell, $\hat {\bf r}={\bf
r}/|{\bf r}|$, and ${\mathcal R}^2$ contains an integral over  radial $2p$ and $3d$ wavefunctions.
One may check that ${\bf D_{\bf k}}^{\dagger}={\bf D_{-\bf k}}$. At this stage, a strong $2p$
spin-orbit coupling can be taken into account by the following replacement
\begin{equation}
p_{{\bf l}m_p\sigma}=\sum\limits_{j,m_j}C^{jm_j}_{1m_p\frac{1}{2}\sigma}p_{{\bf l}jm_j},\label{a7}
\end{equation}
where $C^{jm_j}_{1m_p\frac{1}{2}\sigma}$ are  Clebsch-Gordan coefficients for an electron with a
total angular momentum $j$= 3/2, 1/2. The insertion of Eq.(\ref{a7}) into Eq.(\ref{a6}) splits the
dipole operator: ${\bf D_{\bf k}}={\bf D_{\bf k}}(j=3/2)+{\bf D_{\bf k}}(j=1/2)$. Below we focus on
the $L_3$-edge scattering processes only and, therefore, in Eq.(\ref{a7}) the sum is bounded to
$m_j$ = $\pm 3/2$, $\pm 1/2$,  for $j$ = 3/2.

By applying the circular components $q (=0,\pm 1)$ to vectors $\mbox{\boldmath $\epsilon $}$ and
${\hat {\bf r}}$, the scattering amplitude Eq.(\ref{a2}) can be rewritten as
\begin{equation}
\mathcal F^{\epsilon'\epsilon}_{fg}({\bf k'}, {\bf k}; z_k)=\ \sum\limits_{q',q} T_{q
'q}({\mbox{\boldmath $\epsilon' $},\mbox{\boldmath $\epsilon $}}) F_{q 'q}\left(g\to f;
z_k\right),\label{a8}
\end{equation}
where $T_{q\,'q}({\mbox{\boldmath $\epsilon\,' $},\mbox{\boldmath $\epsilon $}})=
(-1)^{q\,'+q}\epsilon_{-q\,'}'\epsilon_{-q}$ is the polarization tensor and $ F_{q 'q}\left(g\to f;
z_k\right)$ is called the partial amplitude.\cite{Ament11}

From Eqs.(\ref{a2}),(\ref{a6}), and (\ref{a8}) one may see that a scattering process being
considered as the sequence of two local dipole transitions is modulated in space by the phase
factor $\exp[i({\bf k}-{\bf k'}){\bf l}]$, which makes it possible to measure, besides the local
$d$-$d$ excitation, the dispersion laws of accessible propagating electronic
excitations.\cite{Ament11} These include spin and orbital waves, and their mixture in spin-orbital
ordered phases, as well as delocalized excitonic states whose propagation may be largely influenced
by spin and orbital degrees of freedom. Nevertheless, an analysis of the in-gap excitation spectra,
together with the corresponding transition probabilities and polarization dependence, has to be
started with a calculation of a generic local dipole transitions.

The frequently used fast collision approximation\cite{Ament11,Veenendaal96,Brink06,Veenendaal06}
may considerably simplify  a theoretical analysis of RIXS processes. In general outline,  instead
of taking proper account of the individual scattering events that connect the ground state
$|\Psi^N_{[g]}\rangle  $ to a given final state $|\Psi^N_{[f]}\rangle  $ through the intermediary
of core-hole states, in this approximation one simply applies an averaging procedure that consists
in the replacement of the intermediate-state propagator by a resonance factor $G(z_{\bf k})\to
(z_{\bf k}- {\overline \Omega_{res}})^{-1}$. Such an approximation could be well justified if, for
instance,  the lifetime broadening $\Gamma$ of the intermediate core-hole states probed is
comparatively large, which is not the case in vanadium oxides where $\Gamma\approx$ 0.3eV is much
smaller than the intermediate-state interactions.

To go beyond the fast collision approximation one has to take into account the multiplet structure
of the intermediate core-hole states explicitly. The problem is extremely complicated by the fact
that in the transition metal ions involved (i.e., V ions) the intra-atomic 2$p$-3$d$ interactions
are strong, or more precisely, the Coulomb and exchange 2$p$-3$d$ integrals are of the same order
as those for the valence 3$d$ orbitals. As known, the multiplet structure of core-hole states is
directly measured by the X-ray absorption spectroscopy (XAS). Nowadays, XAS spectra for strongly
correlated electronic systems are usually predicted and/or the measured spectra are interpreted
with the use of specially designed computer programs.
\cite{Laan88,Groot90,Ikeno09,Stavitski10,Uldry12}

In principle, the computer code for quantum-chemical cluster calculations we used  for calculating
the multiplet structure of $d^2$ configuration of V$^{3+}$ ion (Section II), can be extended and
applied for solving the core-hole multiplets as well. Nevertheless, it is tempting first to apply
the symmetry-group technique and determine the intermediate-state many-electron wavefunctions of
interest, and then proceed to the calculation of  RIXS amplitudes without using complementary
numerical procedures - the goal that is pursued in the following discussion.

\section{Intermediate core-hole states  in RVO$_3$ }

The first apparent aim is to calculate the on-site matrix elements ( site index ${\bf l}$ is
omitted): $\langle (d^{N+1}\underline{p})_{[i]}|d_{m\sigma}^{\dagger}p_{\frac{3}{2}
m_j}|d^{N}_{[g]}\rangle $ and $\langle d^{N}_{[f]}|p_{\frac{3}{2}
m'_j}^{\dagger}d_{m'\sigma'}|(d^{N+1}\underline{p})_{[i]}\rangle  $, which involve two-electron
states $|d^{N=2}_{[g]/[f]}\rangle  $  of the low-energy configuration $t_{2g}^2$ discussed in
Section II. The intermediate  states $|(d^{N+1}\underline{p})_{[i]}\rangle$ containing one more
electron in 3$d$ valence shell and a core hole in 2$p$ shell have to be determined as eigenvectors
of the local Hamiltonian $H_{loc}$ extended by including  the 2$p$-3$d$ interactions $H_{pd,CX}$
and the spin-orbit coupling $H_{p,so}$ of 2$p$ electrons
\begin{equation}
H_{loc}= H_{cf}+H_{d,C}+ H_{pd,CX}+ H_{p,so}.\label{a9}
\end{equation}
The strength of the direct Coulomb ($C$) and the exchange ($X$) interactions entering the
Hamiltonian $H_{pd,CX}$ are parametrized with the Slater integrals $F^0_{pd}$, $F^2_{pd}$, and
$G^1_{pd}$, $G^3_{pd}$, respectively. For the $3d^{3}2p^5 =(d^{3}\underline{p})$ configuration of V
ion, representative values\cite{Haverkort05} are:  $F^2_{pd} \approx$ 6eV, $G^1_{pd} \approx$ 4.4eV
and $G^3_{pd} \approx$ 2.5eV. The monopole integral $F^0_{pd}$ provides a common shift of all the
multiplets under consideration.

To consider the full multiplet structure of the intermediate core-hole states, an appropriate basis
set is build up in few steps. First, we define the auxiliary states
$|d^3(S_1\Gamma_1)M_1\gamma_1\rangle $ of the $d^3$ configuration not disturbed by $H_{pd,CX}$.
Similar to the $d^2$ configuration, this is done by specifying a set of terms $(S_1\Gamma_1)$, each
composed of degenerate eigenstates of $H_{cf}+H_{d,C}$  with the corresponding eigenvalues
${\mathcal E}_d(S_1\Gamma_1)$. At this step, the approximation can be used that consists in the
neglect in $H_{cf}$ of a weak tetragonal distortion $\Delta \sim$ 0.1eV compared to a strong cubic
CF component $\approx$ 1.5eV. Regarding the intermediate states available in  RIXS processes, this
approximation is justified as follows: Like in X-ray absorption, any CF distortion, as well as a
many-body interaction, that causes small energy effects compared to the lifetime broadening
$\Gamma$ ( $\approx$ 0.3eV for vanadates) cannot be resolved. Loosely speaking, the fast collision
approximation can be safely applied separately to each narrow group of (quasi-)degenerate
intermediate states, but not to the total set of states.

In case of cubic CF symmetry,  the  $t_{2g}^3$ configuration is given by four $(S_1\Gamma_1)$
terms, namely, $^4A_2$, $^2E$, $^2T_1$, and $^2T_2$, with the lowest  $^4A_2$ and  degenerate $^2E$
and $^2T_1$.  With the reference to the lowest level ${\mathcal E}_d(^4A_2)$, the term positions
are ${\mathcal E}_d(^2E)={\mathcal E}_d(^2T_1)=3J_H$ and ${\mathcal E}_d(^2T_2)=5J_H$. For the
further purposes,  four spin states of $^4A_2$  deserve to be presented explicitly (the sole
$\gamma$ is omitted):
\begin{eqnarray}
 &&|t^3(^4A_2), M_1=3/2\rangle   = |d_{\xi \uparrow}d_{\eta\uparrow}d_{\zeta\uparrow}\rangle  , \nonumber\\
 &&|t^3(^4A_2), M_1=1/2\rangle   = \frac{1}{\sqrt{3}}(|d_{\xi \uparrow}d_{\eta\uparrow}d_{\zeta\downarrow}\rangle  +
 |d_{\xi \uparrow}d_{\eta\downarrow}d_{\zeta\uparrow}\rangle  +
 |d_{\xi \downarrow}d_{\eta\uparrow}d_{\zeta\uparrow}\rangle  ), \label{a10}\\
 &&|t^3(^4A_2), M_1=-1/2\rangle   = \frac{1}{\sqrt{3}}(|d_{\xi \downarrow}d_{\eta\downarrow}d_{\zeta\uparrow}\rangle  +
 |d_{\xi \downarrow}d_{\eta\uparrow}d_{\zeta\downarrow}\rangle  +
 |d_{\xi \uparrow}d_{\eta\downarrow}d_{\zeta\downarrow}\rangle  ), \nonumber\\
 &&|t^3(^4A_2), M_1=-3/2\rangle   = |d_{\xi \downarrow}d_{\eta\downarrow}d_{\zeta\downarrow}\rangle \nonumber .
\end{eqnarray}
The states contained in  other terms $(S_1\Gamma_1)$ can be presented in a similar manner as well.

Vector coupling of two irreducible representations $S_1$ and $\Gamma_1$  leads to an equivalent
description of the $t_{2g}^3$ configuration in terms of representations $\overline{\Gamma}$ with
the corresponding states
\begin{equation}
|t^3,\overline{\Gamma}\overline{\gamma}(S_1\Gamma_1)\rangle  =\sum\limits_{M_1,\gamma_1}
|t^3(S_1\Gamma_1)M_1\gamma_1\rangle  U^{S_1\Gamma_1, \overline{\Gamma}}_{M_1\gamma_1,
\overline{\gamma}}\ \label{a11},
\end{equation}
where  coefficients $U^{S_1\Gamma_1, \overline{\Gamma}}_{M_1,\gamma_1, \overline{\gamma}}$ of the
unitary transformation,  Eq. (\ref{a11}), are tabulated\cite{Koster63} (both for the single and
double-valued point groups). Next we note that for the $2p$ core-hole with total angular momentum
$j=3/2$, the states $|\underline{p}_{\frac{3}{2}m_j}\rangle  $ transform according to the
four-dimensional double-valued representation $G_{3/2}$ ($\Gamma_8$ in the Bethe system) of the
cubic group, and the notation $|\underline{p}_{\frac{3}{2}m_j}\rangle =|\underline{p}\ ,
G_{3/2}m_j\rangle$, with $m_j=\pm 3/2, \pm 1/2$, is more convenient.

Finally, the complete basis set $|t^3 \underline{p}\ , \Gamma_J\gamma_J \rangle$ for the core-hole
configuration $t^3\underline {p}$ is determined by  vector coupling of the representations
$\overline{\Gamma}$ for the "undisturbed" valence $t^3_{2g}$ configuration, Eq. (\ref{a11}), and
$G_{3/2}$ for the $2p$ core hole. The coupling relation reads
\begin{equation}
|t^3 \underline{p}\ , \Gamma_J\gamma_J (\overline{\Gamma}G_{3/2}) (S_1\Gamma_1)\rangle
=\sum\limits_{\overline{\gamma},m_j} |t^3, \overline{\Gamma} \overline{\gamma}
 (S_1\Gamma_1)\rangle   |\underline{p}\ , G_{3/2} m_j\rangle  U^{\overline{\Gamma} G_{3/2},
 \Gamma_J}_{\overline{\gamma} m_j, \gamma_J}, \label{a13}
\end{equation}
with the coupling coefficients $U^{\overline{\Gamma} G_{3/2},
 \Gamma_J}_{\overline{\gamma} m_j, \gamma_J}$  defined as in Ref.\onlinecite{Koster63}.
In general, when the procedure outlined above is applied, several repeating representations
  $\Gamma_J, \Gamma_J'$, $\Gamma_J''$, etc., of the same
dimensionality may be generated. Repeating representations are  discriminated with additional
labeling, $(S_1\Gamma_1)$ and $(\overline{\Gamma}G_{3/2})$, kept in Eq.(\ref{a13}), to indicate
different routes $(S_1\Gamma_1)\to(\overline{\Gamma}G_{3/2})\to \Gamma_J$ the representations
$\Gamma_J$ or $\Gamma_J'$, $\Gamma_J''$ arise. Below we  refer to $(S_1\Gamma_1)$ as the generating
multiplet.

 Eq.(\ref{a13}) suggests a conventional basis for a diagonalization of the
local Hamiltonian, Eq.(\ref{a9}). Insofar as states of repeating representations are mixed by
$H_{pd,CX}$, the matrix of $H_{loc}$ takes a block-diagonal form, which can be solved numerically
to give the full multiplet structure of the core-hole configuration $t^3\underline {p}$. Here, the
main difficulty stems from the fact that  complicated vector-coupled functions, Eqs.(\ref{a11}) and
(\ref{a13}), come into play and  calculation of numerous off-diagonal matrix elements of $H_{loc}$
becomes very tedious. Nevertheless, close analysis shows that in several cases calculations are
greatly simplified and some prominent features of the multiplet structure of a core-hole
configuration  can be evaluated without dealing with the full form of  high-rank matrices  of
$H_{loc}$. A particular example is considered in the next Section.

\section{ RIXS transitions in RVO$_3$ }
Let us first apply the outlined procedure for examining the states of those $\Gamma_J$'s that are
related to the lowest generating term $(S_1\Gamma_1)$=$^4A_2$. Because $A_2$ the is one-dimensional
representation, the Eq.(\ref{a11}) allows the following identification
\begin{equation}
|t^3,\overline{\Gamma}\overline{\gamma}(^4A_2)\rangle
=|t^3,\overline{\Gamma}=G_{3/2},\overline{\gamma}= M_1(^4A_2)\rangle  \equiv |t^3(^4A_2),M_1\rangle
, \label{a14}
\end{equation}
with the final kets defined in Eq.(\ref{a10}).  Then, according to Eq.(\ref{a13}), all irreducible
representations $\Gamma_J$'s generated from the term $^4A_2$   are contained in the direct group
product $G_{3/2}\otimes G_{3/2}$ and given\cite{Koster63} in the following decomposition:
$G_{3/2}\otimes G_{3/2}=A_1 + A_2 + E + T_1 + T'_1 + T_2 + T'_2$. Without $2p$-$3d$ interactions
the terms are degenerate, however, being subjected to $H_{pd,CX}$ they are split into two subsets
in such a way that each subset is coupled by $H_{pd,CX}$ to states of other multiplets in different
manners. To exhibit this splitting, the term energies are preliminary found to the first order in
$H_{pd,CX}$ by calculating the corresponding diagonal matrix elements of the local Hamiltonian,
Eq.(\ref{a9}). This yields three lowest terms, $A_2$, $T_1$ and $T_2$, residing at the same energy
that is taken below as the reference (zero) level. Then, the other four terms are shifted to higher
energies as follows: $A_1$ at $\frac{8}{15}G_{pd}^1+\frac{12}{35}G_{pd}^3$, $E$ and $T'_2$ at
$\frac{4}{15}G_{pd}^1+\frac{6}{35}G_{pd}^3$, and $T'_1$ at
$\frac{17}{30}G_{pd}^1+\frac{2}{7}G_{pd}^3$.

The states of three lowest terms, $A_2$, $T_1$ and $T_2$, the first subset, can be regarded as
"edge" resonant states for several reasons. Firstly, to any order in $H_{pd,CX}$ the "edge" states
are not mixed with  other states of the $t^3\underline {p}$ configuration - the corresponding
off-diagonal matrix elements of $H_{loc}$ are checked to be zero. To the contrary, the second
subset, i.e., the states of complementary  multiplets $A_1$, $E$, $T'_2$ and $T'_1$, are mixed up
with states $|t^3 \underline{p}\ , \Gamma_J\gamma_J (S_1\Gamma_1)\rangle$ of similar multiplets
generated from $(S_1\Gamma_1)\not=$$^4A_2$ and located well above the reference level. In total,
this mixture gives rise to a broad and dense manifold of intermediate states $|t^3 \underline{p}\ ,
\Gamma_J\gamma_J\rangle$, but still the energy of "edge" states is retained below the bottom of
this broad manifold. The "edge" states are stabilized mainly due to, first, that the initial energy
of the generating term $^4A_2$ is well below the positions of other terms
$(S_1\Gamma_1)\not=$$^4A_2$ and, second, for $\Gamma_J= A_2$, $T_1$ and $T_2$ the exchange coupling
of the  valence states $|t^3, M_1 (^4A_2)\rangle$ to the core states $|\underline{p}, G_{3/2}
m_j\rangle $ due to $H_{pd,CX}$ is optimized. Experimentally, concerning the $L_3$-edge XAS/RIXS
measurements in RVO$_3$, the "edge" states should be associated with the lowest peak expected at
the very edge of the XAS absorption spectra, while in RIXS  they are accessible by tuning  the
incident photons to the lowest resonance energy $\Omega_{res}^{(1)}$, i.e., as $\hbar\omega_{\bf
k}\to \Omega_{res}^{(1)}$.

Thus emerging multiplet structure of the core-hole configuration $t^3\underline{p}$  suggests  that
close to the "edge"-states associated with the resonance at $\Omega_{res}^{(1)}$, the remaining
terms $\Gamma_J$  evolve into a  peak-shaped structure of several  features at higher energies,
$\Omega_{res}^{(n)} > \Omega_{res}^{(1)}$, for $n=2,3,...$. Particular peaks initially assigned to
different $\Gamma_J$'s are broadened due the core-hole lifetime effects. A group of  peaks may
overlap if their energy spacing is less than the core-hole lifetime broadening.  This means that a
set of core-hole states $|t^3 \underline{p}\ , \Gamma_J\gamma_J \rangle$ that are included  in a
group of "closely spaced" $\Gamma_J$'s, are not resolved and should be attributed to a peculiar
resonance energy $\Omega_{res}^{(n)}$, which is denoted below as $\{\Gamma_J\gamma_J\} {\in (n)}$.
Then the partial amplitude of corresponding local transitions can be expressed in the following
form
\begin{equation}
F_{q 'q}\left(g\to f; \Omega_{res}^{(n)}\right)=\sum\limits_{\{\Gamma_J\gamma_J\}\in
(n)}\frac{\langle t^2,[f]|{\overline {\mathcal D}_{q\prime}}|t^3 \underline{p}\
,\Gamma_J\gamma_J\rangle   \langle t^3 \underline{p}\ ,\Gamma_J\gamma_J|{\mathcal
D}_{q}|t^2,[g]\rangle }{z_k - \hbar\Omega_{res}^{(n)}},\label{a15}
\end{equation}
where ${\mathcal D}_{q}$ is the $q-$th component of the dipole operator, Eq.(\ref{a6}), taken at
${\bf k}=0$. Eq.(\ref{a15}) shows that the states  involved in ${\{\Gamma_J\gamma_J\}\in (n)}$
interfere to yield a joint transition at the resonance energy $\hbar\Omega_{res}^{(n)}$.

To predict the relative positions of $\Omega_{res}^{(n)}$, the  eigenvalue problem should be solved
by diagonalizing numerically the local Hamiltonian, Eq.(\ref{a9}), with the basis set defined in
Eq.(\ref{a13}). Together with the calculated eigenstates $|t^3 \underline{p}\ ;
\Gamma_J\gamma_J\rangle$, this would give one a firm ground for a comprehensive account of RIXS
transition probabilities. As mentioned above, instead of performing this ambitious program, here we
focus on solving a restricted problem - calculation of the RIXS transitions that occur via the
resonant "edge" states, which can be accomplished analytically to the end.

The  explicit form of "edge" states follows directly from Eqs.(\ref{a11}) and (\ref{a13}). In
Eq.(\ref{a13}) the coupling coefficients $U^{\overline{\Gamma} G_{3/2},
\Gamma_J}_{\overline{\gamma} m_j, \gamma_J}$ required for $\overline{\Gamma}=G_{3/2}$ and
$\Gamma_J= A_2$, $T_1$ and $T_2$,  are found from tables of Ref.\onlinecite{Koster63}. For the sake
of brevity, we reduce slightly the notation as follows: $U^{G_{3/2} G_{3/2}, \Gamma_J}_{M_1 m_j,
\gamma_J}\equiv {\mathcal U}^{\Gamma_J \gamma_J}_{M_1 m_j}$. Complementary reduction includes both
$|\underline{p}, G_{3/2}m_j\rangle  =|\underline{p}, m_j\rangle  $, where $m_j=\pm 3/2, \pm 1/2$
for the core-hole states, and $|t^3 \underline{p}; \Gamma_J\gamma_J (G_{3/2} G_{3/2})
(^4A_2)\rangle  = |t^3 \underline{p}; \Gamma_J\gamma_J\rangle  $ for the "edge"-state kets. In
shortened notations the latter can be now expressed as
\begin{equation}
|t^3 \underline{p}\ , \Gamma_J\gamma_J\rangle  =\sum\limits_{M_1,m_j} |t^3,(^4A_2) M_1 \rangle
|\underline{p}, m_j\rangle  {\mathcal U}^{\Gamma_J \gamma_J}_{M_1  m_j}, \label{a16}
\end{equation}
where the kets $|t^3,(^4A_2) M_1 \rangle$ are defined in Eq.(\ref{a10}).

As $\omega_{\bf k}\to \Omega_{res}^{(1)}$, the local RIXS transitions  are described by partial
amplitude, Eq.(\ref{a15}), where $\{\Gamma_J\gamma_J\} {\in (1)}$ denotes  the lowest energy
degenerate multiplets $\Gamma_J= A_2$, $T_1$ and $T_2$ of the $t^3\underline{p}$ configuration.
Matrix elements of the dipole operator are given by
\begin{equation}
\langle t^2,[f]|{\overline {\mathcal D}_{q\prime}}|t^3 \underline{p}\ ,\Gamma_J\gamma_J\rangle
={\mathcal R}^2\sum\limits_{M_1,\sigma}\sum\limits_{m, m_p}\langle
t^2,[f]|d_{m\sigma}|t^3,M_1\rangle \left[\sum\limits_{m_j}{\mathcal U}^{\Gamma_J \gamma_J}_{M_1
m_j}C^{3/2 m_j}_{1m_p\frac{1}{2}\sigma} \right]\langle Y^1_{m_p}|{\hat r}_{q\prime}|Z^2_m\rangle  ,
\label{a17}
\end{equation}
The complex conjugate of Eq.(\ref{a17}) and the replacement $q'\to q$, $[f] \to [g]$ then yields
$\langle t^3 \underline{p}\ , \Gamma_J\gamma_J|{\mathcal D}_{q}|t^2,[g]\rangle $. Having the states
$|t^2,[f]/[g]\rangle$ and $|t^3,M_1\rangle$ as defined in Section II and Eq.(\ref{a10}), together
with the symmetry coupling ${\mathcal U}^{\Gamma_J \gamma_J}_{M_1 m_j}$ and Clebsch-Gordan $C^{3/2
m_j}_{1m_p\frac{1}{2}\sigma}$ coefficients  known,  the transition amplitudes at $\hbar\omega_{\bf
k}\to \Omega_{res}^{(1)}$ are calculated straightforwardly. A connection of the ground state
$|t^2,[g]\rangle$ to particular final states   $|t^2,[f]\rangle$ depicted in Fig.1 and a
polarization dependence of the transitions are of the main interest.

The calculations show that  besides the elastic transition, $|t^2,(^3E)M_ge_g\rangle \to
|t^2,(^3E)M'_ge'_g\rangle$, the only excited term of $t_{2g}^2$ configuration accessible via the
resonance at $\Omega_{res}^{(1)}$ is $^3A_2$, Fig.1. This occurs in the processes
$|t^2,(^3E)M_ge_g\rangle \to |t^2, (^3A_2)M_f\rangle$ with the initial and final states being
degenerate in $M_g(=0, \pm1)$, $e_g$($=e_1,e_2$) and $M_f(=0, \pm1)$, respectively, for the
disordered high-$T$ phase.  As seen from Fig.1,  the corresponding excitation energy is
$\hbar\omega_{fg}=\Delta_t$, and thus the basic CF parameter can be measured directly together with
a fully identified excitonic state. A possible band broadening of this excitonic state is discussed
later on. Identification of this peculiar $d$-$d$ transition in RIXS measurements of RVO$_3$ would
enhance a better understanding of the system. For instance, a trend of interest that $\Delta_t$
develops upon replacement of $R$ ion in RVO$_3$ can be followed. Second, from
Eqs.(\ref{a15}),(\ref{a17}) one obtains  the transition intensity
\begin{eqnarray}
 {\mathcal I}^{\epsilon'\epsilon}(\omega_{fg}\simeq \Delta_t , \Omega_{res}^{(1)}) &=& {\mathcal A}\sum_{M_f}\sum_{M_g,e_g}
\sum\limits_{q',q} |T_{q 'q}({\mbox{\boldmath $\epsilon' $},\mbox{\boldmath $\epsilon $}}) F_{q
'q}\left( (^3E)M_ge_g\to (^3A_2)M_f; \Omega_{res}^{(1)}\right)|^2 ,\label{a18}
\end{eqnarray}
where ${\mathcal A}$ is the known constant.\cite{Ament11} Note that in Eq.(\ref{a18}) an averaging
over the initial states and summation over the final states are performed assuming the state
degeneracy mentioned. From this, the calculated polarization dependence is given up to a factor as
\begin{eqnarray}
{\mathcal I}^{\epsilon'\epsilon}(\omega_{fg}\simeq \Delta_t , \Omega_{res}^{(1)})
&\sim&\left(\epsilon_x^{\,\prime^2}\, +\,\epsilon_y^{\,\prime^2} \right)\epsilon_z^2, \label{a19}
\end{eqnarray}
which may serve for a verification of the previously discussed orbital occupancy both in the
initial and the excited states.

The RIXS $d$-$d$ transitions to the states  of higher-energy  terms $(S_f\Gamma_f)\not=$$^3A_2$
listed in Fig.1 can be measured via the  resonance processes that occur as $\hbar\omega_{\bf k}\to
\Omega_{res}^{(n)}$, with $n>1$. Contrary to the lowest resonance, $n=1$, now the process
$|t^2,(^3E)M_ge_g\rangle \to |t^2,(S_f\Gamma_f)M_f\gamma_f\rangle$ branches off,  meaning that for
given $n(> 1)$ there are several accessible final terms $(S_f\Gamma_f)$. Theoretically, the
intertwined transitions involved could be  unraveled provided the relative energy positions of the
resonance energies $\Omega_{res}^{(n)}$ and the explicit form of intermediate core-hole states
${\Gamma_J\gamma_J\in (n)}$ are found, which would again require using computer codes. Although the
overall inspection of the RIXS processes is not given herein, nevertheless a valuable piece of
information regarding the expected RIXS spectra is predicted, as presented in Fig.1 together with
complementary estimates in Section II, and the lowest $d$-$d$ excitation is studied in great
detail.

So far, when classifying the multiplets of the $t^2$ configurations based on the symmetry-group
technique, we assumed that in high-$T$ phase a deviation from the $D_{4h}$  symmetry is too weak to
be resolved experimentally. With decreasing  temperature, one expects that the deviation increases
and the experimental resolution permits the main effects of the symmetry reduction $D_{4h}\to
D_{2h}$ to be observed in RIXS spectra. Although, a reduction of the former multiplet structure of
$t_{2g}^2$ configuration to a new one due to the local symmetry lowering is a routine theoretical
procedure, a precise analysis of the expected {\it fine} structure of RIXS spectra at low
temperatures is further complicated by the spin-orbital ordering. Actually, since in the lattice
the neighboring V ions are now coupled by the inter-site spin-orbital superexchange, the states of
a local multiplet are split by internal exchange fields and the former degeneracy is removed. In
particular, the symmetry breaking of the ground-state multiplet $^3E$ gives rise to a number of
low-energy spin/orbital excitations whose energy position (and dispersion properties) are set by
the superexchange constant $J\sim$10meV, and hence the excitations are
expected\cite{Miyasaka06,Ishihara04,Sugai06,Jandl10} well below 0.1eV. From the  perspective of
RIXS measurements, the energy region of interest, $\hbar\omega_{fg}< $0.1eV, is typically submerged
by the tail of the huge elastic peak, which makes the selection of the lowest spectral features a
problematic procedure.

In this respect, the promising dispersive quasiparticles for the RIXS measurement originate from
the $d$-$d$ excitations thus far  regarded to be  localized excitons, i.e., the bound electron-hole
pairs in the valence $t_{2g}$ subshell of V ion. A mechanism behind the exciton delocalization is
very similar to that proposed for the orbital wave propagation.\cite{Oles07} It involves a
simultaneous exchange of two {\it active} $t_{2g}$ orbitals at neighboring V-V pairs  forming three
distinguished bonds along lattice axes. For instance, since in the lowest $d$-$d$ excitation of
$^3A_2$ symmetry the $d_\zeta$ orbital involved is {\it inactive} along z$||\bf c$ axis, the
dispersion law of the corresponding exciton is of pure two-dimensional form: $E({\bf q})= E_0
+(W/4)(\cos q_x + \cos q_y)$. Here,  rough estimate for the bandwidth  is $W=\alpha W'$, where
$W'\sim 50$meV and the parameter $\alpha (<1)$ takes into account the temperature dependent spin
and orbital intersite correlations in different phases. The excitonic gap $E_0 $ is determined by
the $t_{2g}$ orbital CF splitting $\tilde\Delta_t$ renormalized compared to $\Delta_t$ due to
intersite superexchange coupling. In disordered high-$T$ phase of RVO$_3$, one has $E_0 =
\Delta_t$, assuming that a small self-energy correction due to virtual intersite
$t^2_{2g}$-electron hoppings is already included into $\Delta_t$ that should be measured. We
conclude this short discussion with a general remark that much more work is needed to discriminate
all possible delocalization mechanisms responsible for a formation of propagating excitonic states.
It seems that a further improvement of the experimental RIXS resolution towards the 10 meV level is
required to measure these quasiparticles.

\section{Conclusion}
To summarize, we developed the RIXS theory and applied it to predict most prominent features of the
low-energy $d$-$d$ excitations that can be measured in RVO$_3$ - a prototypical Mott-Hubbard
insulator with $t^2_{2g}$ configuration  of V$^{3+}$ ions. Complementary {\it ab initio} cluster
calculations allowed us to estimate the crystal field parameters and choose a proper basis set for
vanadium $3d$ orbitals. Initial (ground), intermediate, and final states accessible in RIXS
processes were classified with the use of the symmetry-group technique, which is prerequisite to
the subsequent analysis developed. To go beyond the fast collision approximation frequently used
for calculations of the RIXS transition amplitude, we examined the intermediate core-hole states,
including the spectral and  symmetry properties of many-electron wavefunctions, with more detail
than usually done before. Eventually, the problem of deducing a detailed overall picture for the
intermediate states was reduced to a diagonalization of a local Hamiltonian which involves the
valence electron $d$-$d$ and the core-hole-valence $p-d$ interactions. Mostly because of the huge
number of many-body core-hole states entangled, the diagonalization procedure is far from being a
trivial one.  As well known, such a difficulty one encounters  in any theoretical analysis of XAS
spectra\cite{Laan06} in strongly correlated electron systems, which requires using specially
designed computer codes. At this stage, we avoided doing any numerical calculations, but instead
pursued an analytical treatment so far as possible and found, for instance, that a comprehensive
analysis of the lowest $d$-$d$ excitations is attainable in a simple fashion. For the basic
$t_{2g}^2$ configuration, the relative spectral position of $d$-$d$ excitations of higher energy is
predicted and expressed in terms of a model Hamiltonian describing the local electronic properties
of RVO$_3$. In the present study we restrict ourselves to the calculation of RIXS spectra expected
in a spin- and orbital-disordered high-$T$ phase of RVO$_3$. An extension  to ordered phases, which
is only shortly discussed,  may be a subject of a further work. Delocalization of low-energy
$d$-$d$ excitations leading to a formation of propagating excitonic states was also discussed and
some predictions were formulated. Along this way, a future careful theoretical study and subsequent
examination by RIXS of low-energy excitations in RVO$_3$ will provide a better understanding of the
microscopic mechanisms responsible for rich properties observed in this promising system.

To conclude, we believe that the suggested theoretical approach  may be extended and applied to a
description of selective RIXS transitions in other transition metal oxides with partially filled
$t_{2g}$ orbitals. In further perspective, the {\it ab initio} quantum-chemical cluster approach
can be also applied first to calculate multiplet structure of both the ground-state $t_{2g}^N$ and
the core-hole  $t_{2g}^{N+1}\underline{p}$ configurations and, second, to evaluate the matrix
elements of the dipole transition operator between them. In this context, a symmetry group analysis
developed here for a particular case of V$^{3+}$ can be immediately developed  to provide a
complementary tool for classifying the calculated many-electron wavefunctions and controlling the
symmetry selections of different $d$-$d$ transitions.

\section{Acknowledgments}
We thank J. van den Brink, M. Gr\"uninger, P. Fulde, and L. H. Tjeng  for useful discussions. One
of the authors (V.Yu.) acknowledges partial financial support from the Heisenberg-Landau program.

\newpage
\begin{figure}[b!]
\includegraphics[width=0.85\columnwidth]{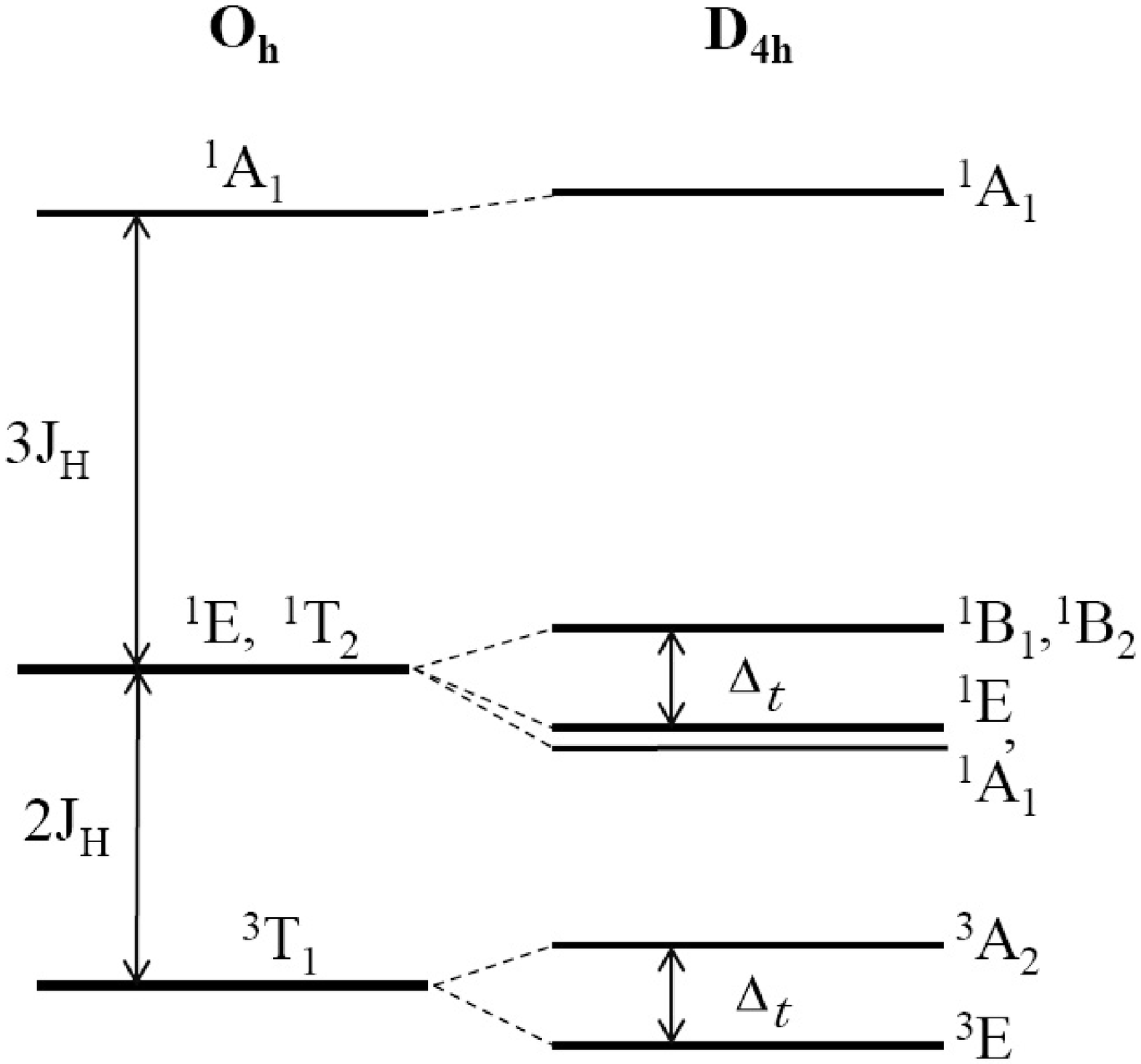}%{Figure-YVo3-t2conf.eps}
\caption { Multiplet structure of $t_{2g}^2$ configuration of
V$^{3+}$ ion in crystal field of
 $D_{4h}$ symmetry. The term branching with the symmetry reduction $O_h \to D_{4h}$ is shown for clarity.
 In RIXS the initial (ground) state is given by $^3E$ and other terms correspond to local
 $d-d$ excitations. More precise estimates for the excitation energies are given in the text.} \label{distortions}
\end{figure}
\end{document}